\documentclass[prd,superscriptaddress,nofootinbib]{revtex4}
\pdfoutput=1
\usepackage{etex}
\reserveinserts{28}

\usepackage{hyperref}
\hypersetup{pdftitle={Drell-Yan Lepton-Pair-Hadron Correlation in pA collisions},pdfauthor={Anna Sta\'sto, Bo-Wen Xiao, David Zaslavsky}}

\usepackage{graphicx}
\usepackage{siunitx}
\usepackage[boldvectors,particle,braket]{physymb}
\usepackage{tikz}
\usepackage{pgfplots}
\usepackage{floatrow}

\usetikzlibrary{backgrounds}
\usetikzlibrary{calc}
\usetikzlibrary{decorations.pathmorphing}
\usetikzlibrary{decorations.markings}
\usetikzlibrary{shapes.geometric}
\usepgfplotslibrary{external}
\tikzset{/tikz/external/mode=list and make}
\tikzset{quark/.style={draw,postaction={decorate},decoration={markings,mark=at position .55 with {\arrow{>}}}}}
\tikzset{antiquark/.style={draw,postaction={decorate},decoration={markings,mark=at position .55 with {\arrow{<}}}}}
\tikzset{lepton/.style={draw,postaction={decorate},decoration={markings,mark=at position .55 with {\arrow{>}}}}}
\tikzset{antilepton/.style={draw,postaction={decorate},decoration={markings,mark=at position .55 with {\arrow{<}}}}}
\tikzset{gluon/.style={decorate,decoration={coil,amplitude=3pt,segment length=5pt}}}
\tikzset{small gluon/.style={decorate,decoration={coil,amplitude=1pt,segment length=2pt}}}
\tikzset{photon/.style={decorate,decoration={snake,amplitude=2pt,segment length=5pt}}}
\tikzset{weak/.style={decorate,decoration={zigzag,amplitude=3pt,segment length=5pt}}}
\newlength\baryonlinespacing
\tikzset{baryon/.style={to path={-- (\tikztotarget) \tikztonodes},
  execute at begin to={
   \draw ($(\tikztostart)!\baryonlinespacing!90:(\tikztotarget)$)--($(\tikztotarget)!\baryonlinespacing!-90:(\tikztostart)$);
   \draw[quark] (\tikztostart)--(\tikztotarget);
   \draw ($(\tikztostart)!\baryonlinespacing!-90:(\tikztotarget)$)--($(\tikztotarget)!\baryonlinespacing!90:(\tikztostart)$);
  }}
 }
\tikzset{right half baryon/.style={to path={-- (\tikztotarget) \tikztonodes},
  execute at begin to={
   \draw[quark] (\tikztostart)--(\tikztotarget);
   \draw ($(\tikztostart)!\baryonlinespacing!-90:(\tikztotarget)$)--($(\tikztotarget)!\baryonlinespacing!90:(\tikztostart)$);
  }}
 }
\tikzset{left half baryon/.style={to path={-- (\tikztotarget) \tikztonodes},
  execute at begin to={
   \draw ($(\tikztostart)!\baryonlinespacing!90:(\tikztotarget)$)--($(\tikztotarget)!\baryonlinespacing!-90:(\tikztostart)$);
   \draw[quark] (\tikztostart)--(\tikztotarget);
  }}
 }
\tikzset{nucleus/.style={to path={-- (\tikztotarget) \tikztonodes},
  execute at begin to={
   \draw ($(\tikztostart)!1.5\baryonlinespacing!90:(\tikztotarget)$)--($(\tikztotarget)!1.5\baryonlinespacing!-90:(\tikztostart)$);
   \draw ($(\tikztostart)!1.2\baryonlinespacing!90:(\tikztotarget)$)--($(\tikztotarget)!1.2\baryonlinespacing!-90:(\tikztostart)$);
   \draw ($(\tikztostart)!0.9\baryonlinespacing!90:(\tikztotarget)$)--($(\tikztotarget)!0.9\baryonlinespacing!-90:(\tikztostart)$);
   \draw ($(\tikztostart)!0.3\baryonlinespacing!90:(\tikztotarget)$)--($(\tikztotarget)!0.3\baryonlinespacing!-90:(\tikztostart)$);
   \draw[quark] (\tikztostart)--(\tikztotarget);
   \draw ($(\tikztostart)!0.3\baryonlinespacing!-90:(\tikztotarget)$)--($(\tikztotarget)!0.3\baryonlinespacing!90:(\tikztostart)$);
   \draw ($(\tikztostart)!0.9\baryonlinespacing!-90:(\tikztotarget)$)--($(\tikztotarget)!0.9\baryonlinespacing!90:(\tikztostart)$);
   \draw ($(\tikztostart)!1.2\baryonlinespacing!-90:(\tikztotarget)$)--($(\tikztotarget)!1.2\baryonlinespacing!90:(\tikztostart)$);
   \draw ($(\tikztostart)!1.5\baryonlinespacing!-90:(\tikztotarget)$)--($(\tikztotarget)!1.5\baryonlinespacing!90:(\tikztostart)$);
  }}
 }
\tikzset{blob/.style={ellipse,minimum height=0.8cm,minimum width=0.3cm}}
\pgfplotsset{compat=1.3,tick scale binop=\times,enlarge x limits=false,enlarge y limits=upper}
\pgfplotsset{correlation graph/.style={xtick={0,1.5708,3.14159,4.7123889,6.2831853},xticklabels={$0$,$\frac{\pi}{2}$,$\pi$,$\frac{3\pi}{2}$,$2\pi$},legend style={font=\small},scaled ticks=false,/pgf/number format/fixed,/pgf/number format/precision=3,max space between ticks=60,try min ticks=3,domain=0:2*pi,xlabel=$\Delta\phi$,ylabel=$\sigma_\text{DYF}/\sigma_\text{DYSIF}$,ymin=0,mark size=0.6pt,mark options={mark=*}}}
\pgfplotsset{ktdist correlation graph/.style={correlation graph,mark options=,no markers,stack plots=y,cycle list name=ktdist colors,axis on top,every axis plot/.append style={fill}}}
\pgfplotsset{colormap={ktdist colors}{
 color=(blue)
 color=(violet)
 color=(magenta)
 color=(red)
 color=(orange)
 color=(yellow)
 color=(green)
 color=(cyan)
 color=(blue)
}}
\pgfplotscreateplotcyclelist{ktdist colors}{
 {violet},
 {magenta},
 {red},
 {orange},
 {yellow},
 {green},
 {cyan},
 {blue}
}

\usepgfplotslibrary{groupplots}
\usepgfplotslibrary{units}

\newcommand\alphasbar{\bar\alpha_\text{s}}
\newcommand\alphaem{\alpha_\text{em}}
\newcommand\sigmadyf{\sigma^{pA\to\gamma^*\pizm X}}
\newcommand\sigmadysif{\sigma^{pA\to\gamma^*X}}
\newcommand\dAu{d--Au}
\newcommand\pPb{p--Pb}
\newcommand\dA{d$A$}
\newcommand\pA{p$A$}
\newcommand\pp{pp}
\newcommand\phasespace{\mathcal{P}.\mathcal{S}.}
\newcommand\ptcut{p_{\perp\text{cut}}}

\begin{document}

\title{Drell-Yan Lepton-Pair-Hadron Correlation in pA collisions}
\author{Anna Sta\'sto}
\email{astasto@phys.psu.edu}
\affiliation{Department of Physics, Pennsylvania State University, University Park, PA
16802, USA}
\affiliation{RIKEN BNL Research Center, Building 510A, Brookhaven National Laboratory,
Upton, NY 11973, USA}
\affiliation{H. Niewodnicza\'nski Institute of Nuclear Physics, Polish Academy of Sciences, Krak\'ow, Poland}
\author{Bo-Wen Xiao}
\email{bux10@psu.edu}
\affiliation{Department of Physics, Pennsylvania State University, University Park, PA
16802, USA}
\author{David Zaslavsky}
\email{dzaslavs@phys.psu.edu}
\affiliation{Department of Physics, Pennsylvania State University, University Park, PA
16802, USA}

\begin{abstract}
In this paper, we numerically study the forward correlations between the
lepton-pair and associated hadrons in Drell-Yan process in \pA{}
collisions. Using the
present knowledge of the dipole gluon distribution from the
modified Golec-Biernat-W\"{u}sthoff model and from the solution of
the Balitsky-Kovchegov evolution equation, we are able to compute
and predict the forward correlations between the lepton-pair and
associated hadron in Drell-Yan process at RHIC and LHC. Similar to
the forward dihadron correlation in \dAu{} collisions measured at
RHIC, the Drell-Yan type correlation also implies a strong
suppression of the away side hadron at forward rapidity due to
the strong interaction between the forward quark from the projectile
proton and the  gluon  density from the target nucleus. Another
feature of this process is that the correlation contains a
double-peak structure in the away side, which makes it a unique
observable.
\end{abstract}

\maketitle

\section{Introduction}
Over many years of both theoretical and experimental studies,
Drell-Yan lepton pair production has been considered as one of the most
interesting processes in high energy physics. Especially, single
inclusive Drell-Yan lepton pair production in \pA{} collisions is
a cleaner probe than high-$p_\perp$ hadrons since it has neither
final state interactions nor fragmentation effects. Therefore, it
provides a unique  opportunity to study the parton
distributions, especially the unintegrated gluon distributions \cite{Catani:1990eg, Collins:1991ty, Avsar:2012hj}, in hadrons. Furthermore, according to the high
energy factorization, the hard photon-hadron and Drell-Yan lepton
pair-hadron (virtual photon-hadron) correlations in \pA{} collisions
serve an important role in probing the low-$x$ dipole gluon
distributions.

In the small-$x$ regime \cite{Iancu:2003xm}, there exist two
distinct unintegrated gluon distributions, namely the
Weizs\"{a}cker-Williams (WW) gluon distribution \cite{Kovchegov:1998bi,McLerran:1998nk} and the dipole
gluon distribution \cite{Kharzeev:2003wz}. According to its operator definition, the WW
gluon distribution corresponds to the conventional gluon
distribution, which measures the gluon density in the light-cone
gauge. Although the dipole gluon distribution has no such  interpretation, it is a fundamental quantity
which  appears in many processes involving hadrons. The dijet (or
dihadron) correlations in DIS \cite{Dominguez:2010xd,Dominguez:2011wm,Dominguez:2011br} can help us
directly measure the WW gluon distribution which has never been
explored experimentally before, while the Drell-Yan
lepton-pair-hadron correlation can directly probe the dipole gluon
distribution at small-$x$.
Since the virtual photon does not interact with the
gluons inside target hadrons,  the final state effects are absent.
As a result,
this correlation can be calculated
exactly in the leading order for all angles between the virtual
photon and associated jet or hadron \cite{Dominguez:2011br}. In contrast, due to the final
state interactions, the so-called correlation limit or
back-to-back limit has to be taken in order to arrive at a
factorized formula for the  dijet or dihadron processes.

Previous studies \cite{Collins:1984kg, Kopeliovich:2000fb,
Baier:2004tj, Gelis:2006hy, Gelis:2002fw,
Gelis:2002nn,GolecBiernat:2010de} of the Drell-Yan processes in
\pp{} or \pA{} collisions have focused primarily on the inclusive
cross sections, but not on the correlations between the outgoing hadron
and the Drell-Yan lepton pair\footnote{We also notice that there
is a recent study\cite{JalilianMarian:2012bd} on the real prompt
photon and hadron correlations in $pA$ collisions. If we take the
virtuality of the photon $M^2$ to be $0$, and subtract the
collinear divergence with proper scheme, we can also get the
results for the real prompt photon and hadron correlations.}. As
shown in Ref.~\cite{Dominguez:2010xd, Dominguez:2011wm,
Dominguez:2011br}, the correlations  can reveal more dynamical
information about the dense nuclear target and yield the direct
measurement of the dipole gluon distribution.

The objective of this paper is to numerically study the
correlations between the lepton-pair and associated hadron in the
Drell-Yan process in \pA{} collisions. This provides predictions
for the correlations which can be measured at both RHIC and
LHC \cite{Salgado:2011wc}.

To parametrize the dipole gluon distribution, we have employed two
approaches. In the first one we used  the Golec-Biernat-W\"{u}sthoff (GBW)
model \cite{GolecBiernat:1998js} together with the geometrical
scaling \cite{Stasto:2000er} which can successfully describe all
the low-$x$ DIS data. In deep inelastic scattering, the partonic
cross section for a virtual photon scattering off a proton can be
written in the color dipole model as a convolution of the photon wave function and the dipole-proton scattering amplitude. In the
GBW model, the dipole scattering amplitude is parametrized as
$S^{(2)}_{\textrm{GBW}}(r_\perp) =\exp \left[-\frac{r_\perp^2
Q_s^2}{4}\right]$ with $r_\perp$ being the transverse size of the
$q\bar q$ dipole. The geometrical scaling property means that
if one writes $Q_s^2(x)= Q_{s0}^2 (x/x_0)^{-\lambda}$ with
$Q_{s0}=\SI{1}{\text{GeV}}$, $x_0=3.04 \times 10^{-3} $ and
$\lambda=0.288$ \cite{Marquet:2006jb}, the DIS total cross section only depends on one
single variable $\tau = \frac{Q^2}{Q_s^2(x)}$, instead of
the variables $x$ and $Q^2$ separately.
Although the geometric scaling was originally found in DIS on
proton targets, it can be generalized to \pA{} collisions as well  \cite{Kowalski:2007rw}. One can write the saturation momentum scale $Q_{sA}$ for
target nuclei with mass number $A$ as $Q_{sA}^2(x) = Q_{s0}^2 A^{\frac{1}{3}} c(b) (x/x_0)^{-\lambda}$, where $c(b)$ is the profile
function which depends on the impact parameter $b$. This profile function is related to the centrality of
the \pA{} (or \dA) collisions. Central collisions give large values
of $c(b)$, while the peripheral collisions correspond to small values of the profile
function.

In the  second approach, we numerically solve the small-$x$ evolution equation for the dipole amplitude, namely the Balitsky-Kovchegov (BK) equation \cite{Balitsky:1995ub+X,Kovchegov:1999yj}, using both fixed and running couplings. From these numerical solutions we then obtain the dipole gluon distribution in each case. We will discuss the implementation of this approach later in more detail.

Comparing to the dihadron (or dijet) correlations in \pA{}
processes, the Drell-Yan type
correlation contains a unique feature on the away side. This correlation has a double-peak structure in the
away side while other correlation functions exhibit a single peak
structure. This difference comes from the fact that the dipole
gluon distribution and the cross section both vanish when the transverse momentum of the
gluon goes to zero. This leads to a vanishing partonic cross section
when the produced virtual photon and quark are completely
back-to-back. With the fragmentation effect which turns the quark
into a pion ($\pi^0$), this effect gets smeared a little bit and
becomes less visible. Nevertheless, taking into account the fragmentation, we find
that the correlation has a minimum at $\Delta \phi=\pi$ and
possesses a two-peak structure.

The rest of the paper is organized as follows. In Sec.~\ref{sec:dijetcorr}, we
discuss the  cross section of the Drell-Yan process with associated hadron
 in \pA{}
collisions and define the correlation function by dividing the single
inclusive cross section of the DY process. We discuss the
parametrization of the dipole gluon distributions in terms of
different models in Sec.~\ref{sec:gluondist}. The numerical results and further
discussions on the future experiments are given in Sec.~\ref{sec:results}. In
Sec.~\ref{sec:conclusion}, we conclude and summarize.

\section{Drell-Yan lepton pair - hadron  Correlations}\label{sec:dijetcorr}
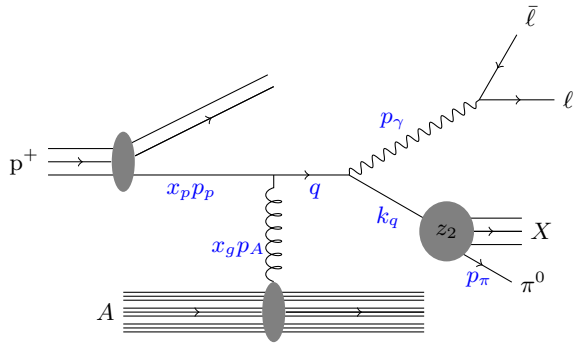
\begin{figure}[tbp]
 \begin{tikzpicture}[momentum/.style={blue}]
  \node[blob] (junction1) at (-2,0) {};
  \node[blob] (junction2) at (0,-2) {};
  \draw[baryon] (junction1) +(-1,0) to (junction1);
  \path (junction1) +(-1,0) node[left] {$\prbr$};
  \draw[left half baryon] (junction1) to +(2,1);
  \draw[quark] (junction1) ++(0,-\baryonlinespacing) node (junction1R) {} -- ++(3,0) node[below,pos=0.3,momentum] {$x_p p_p$} node (junction3) {} node[below,pos=0.85,momentum] {$q$} -- ++(-30:1.5) node[pos=0.4,below,momentum] {$k_q$} node[blob] (junction4) {};
  \draw[quark] (junction4) ++(0,-\baryonlinespacing) -- +(-30:1) node[pos=0.5,below,momentum] {$p_\pi$} node[pos=1,right] {\pizm};
  \draw[baryon] (junction4) to +(1,0);
  \path (junction4) +(1,0) node[right] {$X$};
  \draw[nucleus] (junction2) +(-2,0) to (junction2) to +(2,0);
  \path (junction2) +(-2,0) node[left] {$A$};
  \draw[gluon] (junction2) -- ($(junction1R)!(junction2)!(junction3)$) node[left,pos=0.3,momentum] {$x_g p_A$};
  \draw[photon] (junction3.center) -- +(30:2) node[above left,pos=0.5,momentum] {$p_\gamma$} node (junction5) {};
  \draw[lepton] (junction5.center) -- +(0:1) node[right] {$\ell$};
  \draw[antilepton] (junction5.center) -- +(60:1) node[above right] {$\bar\ell$};
  \node[fill=gray,blob] at (junction1) {};
  \node[fill=gray,blob] at (junction2) {};
  \node[fill=gray,blob] at (junction4) {$z_2$};
 \end{tikzpicture}
 \caption{The Drell-Yan scattering process, with several of the momentum variables used in our calculation labeled. $p_p$ and $p_A$ are respectively the momenta of the proton and the nucleus from which the gluon was emitted. $p_\gamma$ is the momentum of the virtual photon as reconstructed from the lepton pair, and $p_\pi$ is the measured momentum of the pion. $x_p$ and $x_g$ are the longitudinal momentum fractions of the quark relative to the proton and the gluon relative to the nucleon, $z = p_\gamma^+/q^+$ is the fraction of total momentum taken by the photon, and $z_2 = p_\pi^+/k_q^+$ is the fraction of the momentum of the quark jet that is carried by the pion.}
 \label{fig:scattering}
\end{figure}

In the small-$x$ regime, the Drell-Yan lepton pair production
processes is dominated by the  $qg\to q\gamma^*$ channel at the
partonic level. Here the gluon $g$, in fact, can be regarded as a
group of soft gluons from the dense nucleus target. The virtual
photon then decays into a lepton pair and the quark fragments into a
cluster of hadrons. We are interested in the correlation between
the lepton pair and the hadron, say $\pi^0$, which is a product of the fragmentation from the quark
in the process ${p A\to\ell\bar\ell\pizm X}$. It is shown
schematically in figure \ref{fig:scattering}.

The cross section $\ud{\sigmadyf}{\phasespace}$, which measures the probability of producing a virtual photon with an invariant mass $M$ and a quark with large $p_\perp$ in the final state, has been calculated in \cite{Dominguez:2011br} for the $p A\to\gamma^*qX$ channel. After summing over different photon polarizations, the total  production cross section can be cast into\footnote{The rapidity is defined as respect to the center of mass frame of the scattering,  which coincides with the lab frame when the energy of the proton projectile is the same as the energy per nucleon in the nucleus target.}
\begin{multline}
\frac{\udc\sigma^{pA\to\gamma^*qX}}{
\udc Y_\gamma\udc Y_q \uddc \vec{p}_{\gamma\perp} \uddc \vec{k}_{q\perp}\uddc b} = \sum_{f}x_{p}q_{f}(x_{p},\mu)\frac{\alphaem e_f^2}{2\pi^{2}}
 (1-z)  F_{x_g}(q_\perp) \\
\times \Biggl\{ \bigl[1 + (1-z)^2\bigr] \frac{z^2 q_{\perp}^2}{\bigl[\tilde{P}_{\perp}^2+\epsilon_M^2\bigr]\bigl[(\tilde{\vec{P}}_{\perp}+z\vec{q}_\perp)^2+\epsilon_M^2\bigr]} \\
- z^2(1-z) M^2 \Biggl[\frac{1}{\tilde{P}_{\perp}^2 + \epsilon_M^2} - \frac{1}{(\tilde{\vec{P}}_{\perp}+z\vec{q}_\perp)^2 + \epsilon_M^2}\Biggr]^2 \Biggr\},
\end{multline}
where $\epsilon_M^2=(1-z)M^2$, $\vec{q}_\perp=\vec{p}_{\gamma\perp}+\vec{k}_{q\perp}$ and $\tilde{\vec{P}}_{\perp} = (1-z)\vec{p}_{\gamma\perp}-z\vec{k}_{q\perp}$. Here the normalized gluon distribution, $F_{x_g}(q_\perp)=\int \frac{d^2 r_\perp}{(2\pi)^2} e^{-iq_\perp \cdot r_\perp} S^{(2)}_{x_g}(r_\perp)$, is defined through the Fourier transform of the dipole amplitude. It is straightforward to write down the corresponding cross section for the lepton pair production as follows
\begin{equation}
\frac{\udc\sigma^{pA\to l^+l^- qX}}{
\udc Y_\gamma\udc Y_q \uddc \vec{p}_{\gamma\perp} \uddc \vec{k}_{q\perp}\uddc b\, \uddc M^2}=\frac{\alphaem}{3\pi M^2} \frac{\udc\sigma^{pA\to\gamma^*qX}}{%
\udc Y_\gamma\udc Y_q \uddc \vec{p}_{\gamma\perp} \uddc \vec{k}_{q\perp}\uddc b} \; .
\end{equation}
Therefore, one just needs to include a factor of $ \frac{\alphaem}{3\pi} \frac{\udc M^2}{M^2}$ to the photon cross sections to get the lepton pair cross sections. The factor $ \frac{\alphaem}{3\pi} \frac{1}{M^2}$ eventually cancels out in the correlation function \eqref{eq:correlation}.

To express the cross section in terms of the kinematic variables of the pion ($\pi^0$) and leptons, which are actually detected in the final state, we need to include in the above formula the fragmentation function of the quark into a pion, $D_{\pi^0/f}(z_2,\mu)$ to obtain
\begin{multline}\label{eq:exclusivedsigma}
\frac{\udc\sigmadyf}{%
\udc Y_\gamma \udc Y_\pi \uddc \vec{p}_{\gamma\perp} \uddc \vec{p}_{\pi\perp} \uddc b} = \int^1_{\frac{z_{h2}}{1-z_{h1}}}\frac{\udc z_2}{z_2^2} \sum_{f}D_{\pizm/f}(z_2,\mu)x_{p}q_{f}(x_{p},\mu)\frac{\alphaem e_f^2}{2\pi^{2}} (1-z)  F_{x_g}(q_\perp)\\
 \times\Biggl\{ \bigl[1 + (1-z)^2\bigr] \frac{z^2 q_{\perp}^2}{\bigl[\tilde{P}_{\perp}^2+\epsilon_M^2\bigr]\bigl[(\tilde{\vec{P}}_{\perp}+z\vec{q}_\perp)^2+\epsilon_M^2\bigr]} \\
- z^2(1-z) M^2 \Biggl[\frac{1}{\tilde{P}_{\perp}^2 + \epsilon_M^2} - \frac{1}{(\tilde{\vec{P}}_{\perp}+z\vec{q}_\perp)^2 + \epsilon_M^2}\Biggr]^2 \Biggr\} \; ,
\end{multline}
where $z_{h1} \equiv p_\gamma^+/p_p^+$ and $z_{h2} \equiv p_\pi^+/p_p^+$ are the longitudinal momentum fraction of the virtual photon and final state $\pizm$ hadron, respectively. $\mu^2$ is the factorization scale which is set to be $Q_{sA}^2$. The lower limit on the $z_2$ integral comes from requiring that $x_p < 1$.

The Drell-Yan lepton pair inclusive differential cross section can be obtained from the previous expression by integrating over the phase space of the final state quark, which makes it a function of the kinematic variables of the virtual photon only. By noting that $\udc Y_\gamma \udc Y_q =\frac{\udc z \udc x_p}{z(1-z) x_p} $, one can write the inclusive Drell-Yan cross section as
\begin{multline}\label{eq:inclusivedsigma}
\frac{\udc\sigmadysif}{%
\udc Y_\gamma \uddc \vec{p}_{\gamma\perp}\uddc b} = \int_{z_{h1}}^1 \frac{\udc z}{z} \iint \uddc \vec{q}_{\perp}\sum_{f}x_{p}q_{f}(x_{p},\mu)\frac{\alphaem e_f^2}{2\pi^{2}}  F_{x_g}(q_\perp) \\
\times\Biggl\{ \bigl[1 + (1-z)^2\bigr] \frac{z^2 q_{\perp}^2}{\bigl[p_{\gamma\perp}^2+\epsilon_M^2\bigr]\bigl[(\vec{p}_{\gamma\perp}-z\vec{q}_\perp)^2+\epsilon_M^2\bigr]} \\
- z^2(1-z) M^2 \Biggl[\frac{1}{p_{\gamma\perp}^2 + \epsilon_M^2} - \frac{1}{(\vec{p}_{\gamma\perp}-z\vec{q}_\perp)^2 + \epsilon_M^2}\Biggr]^2 \Biggr\}.
\end{multline}
This result is in agreement with the results in Ref.~\cite{Kopeliovich:2000fb,Baier:2004tj,Gelis:2002nn,Gelis:2002fw}.

Integrating \eqref{eq:exclusivedsigma} and
\eqref{eq:inclusivedsigma} over the magnitude of the transverse
momenta with certain cutoff gives the cross sections as  a
function of rapidity and impact parameter only,
$\frac{\udc\sigmadyf}{\udc Y_{\gamma}\udc Y_{\pi}\uddc\vec{b}}$
and $\frac{\udc\sigmadysif}{\udc Y_{\gamma}\uddc\vec{b}}$. To set
the lower limits on these momentum integrals, we impose a
transverse momentum cutoff $\ptcut$ which can be set to match the
experimental acceptance. The upper limits are again
determined by requiring that the longitudinal momentum fraction of
the initial quark $x_p$ be less than 1. This would give the lower
bound of $z_2$ integrals and upper bounds of momentum integrals.
For the correlation measurement, either the virtual photon or the
detected hadron can be the leading particle with larger transverse
mass, while the other becomes the associated particle. In the
code, we break the total contributions into two cases: (1)
$\sqrt{p_{\gamma\perp}^2+M^2} >p_{q\perp}$;
(2)$\sqrt{p_{\gamma\perp}^2+M^2}<p_{q\perp}$ which correspond to
the region $z >\frac{1}{2}$ and  $z <\frac{1}{2}$, respectively.

Finally, the ratio of these two cross sections defines the Drell-Yan pair-hadron correlation function $C^{\text{DY}}(\Delta\phi)$,
\begin{equation}\label{eq:correlation}
 C^{\text{DY}}(\Delta \phi) =\frac{ 2\pi  \displaystyle\idotsint\limits_{p_{\{\gamma,\pi\}\perp} > \ptcut}\udc \vec{p}_{\gamma\perp}\vec{p}_{\gamma\perp} \udc\vec{p}_{\pi\perp}\vec{p}_{\pi\perp}\frac{\udc\sigma^{pA\to\gamma^*\pizm X}}{\udc Y_\gamma \udc Y_\pi \uddc \vec{p}_{\gamma\perp} \uddc \vec{p}_{\pi\perp} \uddc b} }{\displaystyle\iint\limits_{p_{\gamma\perp} > \ptcut}\uddc\vec{p}_{\gamma\perp}\frac{\udc\sigma^{pA\to\gamma^*X}}{\udc Y_\gamma \uddc \vec{p}_{\gamma\perp}\uddc b}},
\end{equation}
where $\Delta \phi$ is defined as the azimuthal angle difference between the virtual photon and measured $\pi^0$. The correlation function expresses the probability density for a pion to be emitted at a specific azimuthal angle relative to the virtual photon (normally the rapidities of these two particles are chosen to be the same). When $\Delta \phi \simeq \pi$, that is, when the virtual photon and $\pi^0$ are almost back-to-back, the dominant contribution to the correlation function $C^{\text{DY}}(\Delta \phi)$ comes from the low $q_\perp$ region of the gluon distribution $F_{x_g}(q_\perp)$. In contrast, the behavior of the correlation function $C^{\text{DY}}(\Delta \phi)$ at $\Delta \phi \simeq 0 \text{ or } 2\pi$ is mainly determined by the large transverse momentum part of the gluon distribution function $F_{x_g}(q_\perp)$.

In addition,as mentioned earlier, the partonic cross section vanishes at $q_\perp =0$ as in Eq.~(\ref{eq:exclusivedsigma}). The physical interpretation is that the emission of the virtual photon from the quark requires that the quark gets a kick or an impulse in the transverse direction. Since the small-$x$ gluons carry very little longitudinal momentum and the virtual photon does not interact with the dense small-$x$ gluons from the target nucleus, the quark must change its transverse momentum during the multiple scattering with the gluons in order to radiate the photon. In other words, if the quark receives no impulse which change its momentum, the virtual photon will remain as part of the quark higher fock state wavefunction, and thus can not be produced in the final state. Due to this fact, we expect that $\Delta \phi= \pi$ should be a local minimum of the correlation function. Furthermore, there are two symmetrical maxima lying near $\Delta \phi= \pi$ as $q_\perp^2 F_{x_g}(q_\perp)$ increases with $q_\perp$ and reaches its maximum at $q_\perp \simeq Q_s$. As a result, the correlation should have a unique double-peak structure on the away side. Note that the correlation function at  $\Delta \phi= \pi$ is no longer zero due to the pion fragmentation function.

In terms of the numerical evaluation of the cross sections \eqref{eq:exclusivedsigma} and \eqref{eq:inclusivedsigma}, we use the MSTW 2008 NLO parton distributions \cite{Martin:2009iq} for the parton distributions $q_f$ and the DSS fragmentation functions \cite{deFlorian:2007aj} for the fragmentation functions $D_{\pizm/q}$.  In the case of the \dAu{} collisions at RHIC, since the parton distributions are only given for protons in Ref.~\cite{Martin:2009iq}, we obtain the parton distribution for neutrons by using the isospin symmetry, which gives $q_{\mathrm{u/p}} = q_{\mathrm{d/n}}$ and $q_{\mathrm{d/p}} = q_{\mathrm{u/n}}$.

\section{The Dipole Gluon Distribution}\label{sec:gluondist}
The Drell-Yan cross section involves the so-called normalized
dipole gluon distribution $F_{x_g}(q_\perp)$, which characterizes
the dense gluon distribution in this process in the nucleus in the
small-$x$ regime. As discussed in Ref.~\cite{Dominguez:2010xd}, this
function is related to the well-known dipole gluon distribution
$xG^{(2)}$ as
\begin{equation}
 xG^{(2)}(x, q_\perp) = \frac{N_c S_\perp}{2\pi^2\alpha_s} q_\perp^2 F_{x_g}(q_\perp) \; ,
\end{equation}
where $S_\perp$ is the transverse area of the target nucleus, $N_c = 3$ is the number of colors, and $\alpha_s$ is the strong coupling constant.

As mentioned previously we have chosen two approaches  to parametrize the normalized dipole gluon
distribution $F_{x_g}(q_\perp)$, namely, the solution of the BK
evolution equation \cite{Balitsky:1995ub+X,Kovchegov:1999yj} and
the GBW model \cite{GolecBiernat:1998js}.
$F_{x_g}$  is a Fourier transform of the color dipole
amplitude in the coordinate space. The implementation of the solution
to the leading order BK equation for the phenomenological use is a
little bit troublesome since it gives too fast energy evolution as
compared to the experimental data from DIS. The GBW model
describes the DIS data very well and is straightforward to use.
However the GBW model fails to describe the proper behavior gluon distribution at large transverse momentum since it has an exponential drop off, while the perturbative QCD gives a result which drops in terms of powers of $q_\perp$. In practice, we use
these two approaches for our numerical evaluation since they are complementary to each other. Our procedure is to fine-tune the behavior of the strong coupling constant $\alpha_s$ or the running coupling $\alpha_s(q)$ in the numerical solution of the BK equation until the solution has the same energy dependence and similar low $q_\perp$ behavior as what the GBW model gives. Then we believe that the solution of the BK equation should be able to produce the correct large $q_\perp$ part of the distribution and thus the correct angle correlation at $\Delta \phi =0, 2\pi$.

\subsection{The BK Equation}

The Balitsky-Kovchegov (BK) equation governs the rapidity
dependence of the color dipole scattering amplitude
$\mathcal{N}(\vec{r}, \vec{b},
Y)$ \cite{Balitsky:1995ub+X,Kovchegov:1999yj}. In position space,
we can write the equation in terms of $S(\vec{r}, \vec{b}, Y) = 1
- \mathcal{N}(\vec{r}, \vec{b}, Y)$, as
\begin{multline}
 \pd{S(\vec{x} - \vec{y}, \vec{b}, Y)}{Y} = -\bar\alpha_s\iint\frac{\uddc\vec{z}}{2\pi}\frac{(\vec{x} - \vec{y})^2}{(\vec{x} - \vec{z})^2(\vec{z} - \vec{y})^2}\biggl[S(\vec{x} - \vec{y}, \vec{b}, Y) \\
  - S\biggl(\vec{x} - \vec{z}, \vec{b} + \frac{\vec{z} - \vec{y}}{2}, Y\biggr)S\biggl(\vec{z} - \vec{y}, \vec{b} + \frac{\vec{x} - \vec{z}}{2}, Y\biggr)\biggr] \;.
\end{multline}
The above form of the equation is valid in the leading logarithmic order in $\ln 1/x$.
One can also write the BK equation in terms of the momentum-space dipole scattering amplitude,
\begin{equation}\label{eq:phifromN}
 \phi(\vec{k}, Y) = \iint\frac{\uddc\vec{r}}{2\pi}e^{-i\vec{k}\cdot\vec{r}}\frac{N(\vec{r} , Y)}{r^2} \; .
\end{equation}

As shown in Ref.~\cite{GolecBiernat:2001if}, in terms of the function $\phi(\vec{k}, Y)$, the BK equation reads
\begin{equation}
 \pd{\phi(k, Y)}{Y} = \bar{\alpha}_s K\otimes\phi(k) - \bar\alpha_s\phi^2(k) \; ,
\end{equation}
where $K$ represents the action of the BFKL kernel, which in the leading logarithmic approximation reads
\begin{equation}
 K\otimes\phi(k) = \int_0^\infty\frac{\udc k'^2}{k'^2} \biggl[\frac{k'^2\phi(k') - k^2\phi(k)}{\abs{k^2 - k'^2}} + \frac{k^2\phi(k)}{\sqrt{4k'^4 + k^4}}\biggr] \; .
\end{equation}
The Laplacian of $\phi$ is the function $F_{x_g}$ that appears in the cross sections of section \ref{sec:dijetcorr}.
\begin{equation}\label{eq:fgfromphi}
 F_{x_g}(\vec{k}, \vec{b}) = \frac{1}{2\pi}\nabla_{\vec{k}}^2 \phi\bigl(\vec{k}, \vec{b}, Y(x_g)\bigr) + \delta^2(\vec{k}) \; .
\end{equation}
Since the partonic cross section vanishes at $k=0$, we can safely
ignore the delta function in the above equation. Beyond this
section, the $\vec{b}$ dependence is left implicit.

The normalized dipole gluon distribution $F_{x_g}$ at small $x$ extracted from the solution to the leading order BK equation yields   energy dependence  which is too fast as compared with phenomenology. In addition, the numerical solution to the full NLO BK equation \cite{Avsar:2011ds} is not yet ready for use in phenomenological study. In practice, we have chosen to  modify the solution from the LO BK equation to obtain a realistic energy dependence by varying the behavior of the coupling constant. For a fixed coupling, we vary the value of the coupling constant directly; for the running coupling, we vary the value of $\Lambda_\text{QCD}$ which controls the running behavior.

\subsection{The Golec-Biernat Wusthoff Model}
We can also parametrize the dipole gluon distributions by using the GBW model adapted for nuclear targets, which has both the nuclear enhancement and geometrical scaling \cite{Stasto:2000er, Iancu:2002tr} feature. The geometrical scaling was found to be related to the traveling wave solutions \cite{Braun:2000wr, Munier:2003vc,Mueller:2002zm} of the Balitsky-Kovchegov evolution.

For a dense proton target, the GBW model reads
\begin{align}
\int \uddc\vec{b}\,\mathcal{N}(\vec{r}, \vec{b}, Y) &= \mathcal{N}_0\bigl(1 - e^{-r^2 Q_s^2/4}\bigr)  \; , & Q_s^2(Y) &= Q_{s0}^2 x_0^\lambda e^{\lambda Y} = Q_{s0}^2 \biggl(\frac{x_0}{x}\biggr)^{\lambda},
\end{align}
where $Q_{s0}^2 = \SI{1}{GeV^2}$, $\lambda=0.288$ and $x_0=3.04\times 10^{-4}$. As for a dense nuclear target with $A$ nucleons inside, the saturation scale is subject to the nuclear enhancement which modifies the saturation momentum and gives $Q_{sA}^2(x) = Q_{s0}^2 A^{\frac{1}{3}} c(b) (x/x_0)^{-\lambda}$. Here the profile function $c(b)$ gives the centrality of the scattering. Using the GBW model, we can compute a simple analytic expression for $F_{x_g}$. According to equations \eqref{eq:phifromN} and \eqref{eq:fgfromphi} above, we get
\begin{align}
 \phi(k^2, Y) &= \frac{1}{2}\Gamma\biggl(0, \frac{k^2}{Q_{sA}^2(Y)}\biggr) \label{eq:gbwphi} \; , \\
 F_{x_g}(k^2, Y) &= \frac{1}{\pi Q_{sA}^2(Y)}e^{-k^2/Q_{sA}^2(Y)} \; ,  \label{eq:gbwfg}
\end{align}
where $\Gamma[0,x]$ is the incomplete gamma function.

\subsection{The numerical solution of the BK equation}

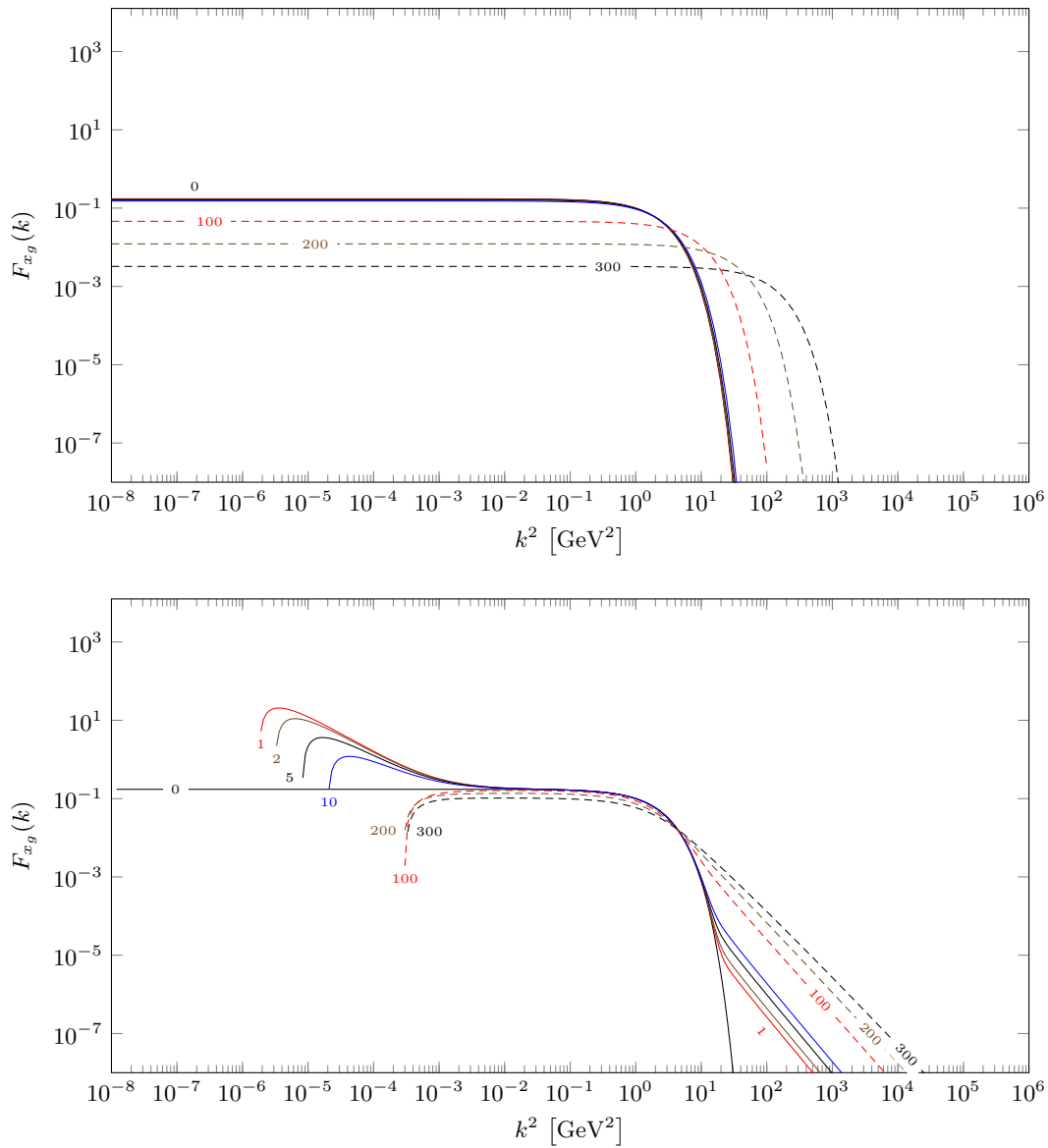
\begin{figure}
 % This figure may require at least PGF 2.10 to compile correctly.
 \begin{tikzpicture}
  \begin{axis}[xmode=log,ymode=log,domain=1e-8:1e2,samples=100,xmin=1e-8,xmax=1e6,ymin=1e-8,ymax=1e3,xlabel=$k^2$,ylabel=$F_{x_g}(k)$,x unit=GeV^2,width=14cm,height=8cm]
   \addplot+[black,no markers] {0.17341*exp(-\x/1.83559)} node[pos=0.04,above,font=\tiny] {$0$};
   \addplot+[no markers] {0.171125*exp(-\x/1.8601)};
   \addplot+[no markers] {0.168871*exp(-\x/1.88493)};
   \addplot+[no markers] {0.162284*exp(-\x/1.96144)};
   \addplot+[no markers] {0.151871*exp(-\x/2.09592)};
   \addplot+[no markers,domain=1e-8:1e2] {0.0460337*exp(-\x/6.91471)} node[pos=0.1,font=\tiny,fill=white] {$100$};
   \addplot+[no markers,domain=1e-8:1e3] {0.0122202*exp(-\x/26.0479)} node[pos=0.12,font=\tiny,fill=white] {$200$};
   \addplot+[no markers,domain=1e-8:1e4] {0.00324398*exp(-\x/98.1231)} node[pos=0.14,font=\tiny,fill=white] {$300$};
  \end{axis}
  \begin{axis}[yshift=-8cm,xmode=log,ymode=log,xmin=1e-8,xmax=1e6,ymin=1e-8,ymax=1e3,xlabel=$k^2$,ylabel=$F_{x_g}(k)$,x unit=GeV^2,width=14cm,height=8cm]
   \addplot+[black,no markers] file {datafiles/Fg/BK-Pb85-000.dat} node[pos=0.04,font=\tiny,fill=white] {$0$};
   \addplot+[no markers] file {datafiles/Fg/BK-Pb85-001.dat} node[pos=0,below,font=\tiny] {$1$} node[pos=0.5,sloped,below,font=\tiny] {$1$};
   \addplot+[no markers] file {datafiles/Fg/BK-Pb85-002.dat} node[pos=0,below,font=\tiny] {$2$};
   \addplot+[no markers] file {datafiles/Fg/BK-Pb85-005.dat} node[pos=0,left,font=\tiny] {$5$};
   \addplot+[no markers] file {datafiles/Fg/BK-Pb85-010.dat} node[pos=0,below,font=\tiny] {$10$};
   \addplot+[no markers] file {datafiles/Fg/BK-Pb85-100.dat} node[pos=0,below,font=\tiny] {$100$} node[pos=0.5,sloped,font=\tiny,fill=white] {$100$};
   \addplot+[no markers] file {datafiles/Fg/BK-Pb85-200.dat} node[pos=0,left,font=\tiny] {$200$} node[pos=0.55,sloped,font=\tiny,fill=white] {$200$};
   \addplot+[no markers] file {datafiles/Fg/BK-Pb85-300.dat} node[pos=0,right,font=\tiny] {$300$} node[pos=0.6,sloped,font=\tiny,fill=white] {$300$};
  \end{axis}
 \end{tikzpicture}
 \caption{The gluon distribution from the GBW model \eqref{eq:gbwfg}, on top, and the output of the numerical BK evolution with fixed coupling \eqref{eq:fgfromphi}, on bottom, for central \pPb{} collisions. Each curve shows $F_{x_g}(k)$ for $x_g = \exp(Y_\text{init} - n\delta Y)$, where $n$ is the number that labels the curve and $\delta Y = \frac{1}{400}\ln(10^8) \approx 0.04605$ is the step size in rapidity used by the numerical integrator. The curve labeled 0 is the initial condition. Note the transient ``spike'' at the low $k^2 $ region of the solution to the BK evolution, and the significant enhancement at large $k^2$ relative to the GBW model.}
 \label{fig:fgplot}
\end{figure}

We solve the BK equation numerically~\cite{GolecBiernat:2001if}  with initial condition for  $ \phi(k^2, Y)$ as in Eq.~\eqref{eq:gbwphi} at $Y_\text{init} = -\ln x_\text{init}$, where we use $x_\text{init} = 0.01$. This value  is chosen to coincide with the upper applicable limit of the GBW model. With a solution for $\phi$, using $F_{x_g}(k^2, Y) = \frac{1}{2\pi k^2}\pdd{\phi(k^2, Y)}{(\ln k)}$, % was there a reason for keeping 1/2pi separate?
we then compute $F_{x_g}$ in the region $Y > Y_\text{init}=4.6$ using a proper numerical differentiation scheme. The numerical solution to the BK equation for the gluon distribution  is shown in figure~\ref{fig:fgplot} with fixed coupling, and in figure~\ref{fig:fgplotrunning} for running coupling, as a function of the transverse momentum for different rapidities.
For comparison also the analytic expression from the GBW model \eqref{eq:gbwfg} is shown in figure~\ref{fig:fgplot}. The BK solution possesses a larger tail in momentum as compared with the GBW model, which is due to the correct perturbative QCD limit.

\subsubsection{Fixed Coupling}

\begin{figure}
 \begin{tikzpicture}[every axis/.append style={ymode=log,xlabel=$Y$,ylabel=$k_\text{max}$,y unit=GeV,
  cycle list={{blue,mark=square*},{brown,mark=diamond*},{black,thick,mark=*},{green!90!black,mark=triangle*},{red,mark=pentagon*}}}]
  \begin{axis}[width=11cm,height=6cm,legend style={at={(1.1,0.5)},anchor=west},legend cell align=left,legend columns=1]
   \foreach \as in {as200,lam0588,GBW,lam0010,as062} {%
    \addplot+[mark repeat=100] file {datafiles/asmatching/\as.peak.dat};
   }
   \legend{$\alpha_s = 0.2$,$\Lambda_\text{QCD}^2 = 0.0588$,GBW,$\Lambda_\text{QCD}^2 = 0.0010$,$\alpha_s = 0.062$}
  \end{axis}
 \end{tikzpicture}
 \caption{This plot shows the peak of the momentum distribution, $k_\text{max} = \max k\phi(k^2, Y)$, computed from the analytic formula for the GBW model, from the fixed coupling BK evolution for selected values of $\alpha_s$, and from the running coupling BK evolution with selected values of $\Lambda_\text{QCD}^2$. For the BK evolution curves, the slope in the upper range of rapidities decreases as $\alpha_s$  or $\Lambda_\text{QCD}^2$ decreases, and the closest match to the slope of the GBW model curve at $Y > 15$ is achieved with $\alpha_s = 0.062$ for fixed coupling or $\Lambda_\text{QCD}^2 = 0.001$ for running coupling. The jagged ``steps'' in the curves reflect the finite spacing of the momentum grid used in the evolution.}
 \label{fig:kmax}
\end{figure}
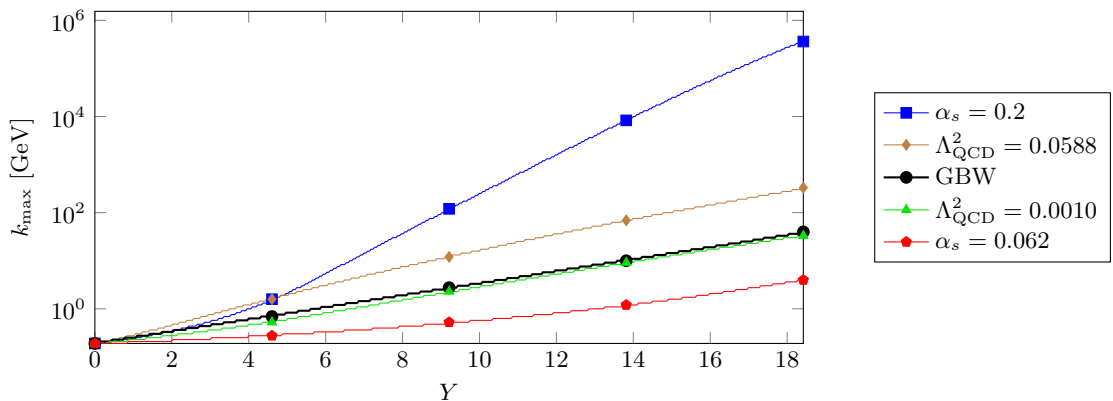

Since the solution to the leading order BK equation gives
 energy dependence which is too fast as compared to data (for inclusive DIS)
if we use the normal value for $\alpha_s$, we treat the strong
coupling constant $\alpha_s$ as an adjustable parameter in the
numerical computation. We need to find out the value of $\alpha_s$
which corresponds to the kinematic regime probed by these
collisions in order to make meaningful predictions. Theoretically, the value of $\lambda$ is predicted by various
analytical approaches to the solution to the BK equation
\cite{Mueller:2002zm,Munier:2003vc}. The resulting saturation scale can be cast into
\begin{equation}\label{eq:theoreticalevolution}
 Q_s^2(Y) = \frac{Q_{s0}^2}{Y^{\frac{3}{2(1-\gamma_0)}}} \exp\biggl(-\frac{\bar\alpha_s \chi(\gamma_0)}{1 - \gamma_0} Y\biggr), % Q_s0 or Q_sA?
\end{equation}
where $\bar \alpha_s =\frac{\alpha_s N_c}{\pi}$, $\chi (\gamma)$ is the usual characteristic function the value of $\gamma_0$ is determined to be around $0.37$.
This function asymptotically approaches an exponential form,
\begin{equation}\label{eq:asymptoticbk}
 Q_s^2(Y) \to C\exp(-\lambda Y),\quad\text{where}\ \lambda = \frac{\bar\alpha_s \chi(\gamma_0)}{1 - \gamma_0} = \frac{4.88 N_c}{\pi}\alpha_s.
\end{equation}

In terms of the numerical computation, as shown in figure \ref{fig:kmax}, we extract the saturation scale from the output of the BK evolution and employ the procedure as described in Ref.~\cite{GolecBiernat:2001if}, determining the value $k_\text{max}$ of $k$ which maximizes $k \phi(k^2, Y)$ at each $Y$, and set the saturation scale to be proportional to $k_\text{max}$. It is straightforward to show that the assumption $k_\text{max} \propto Q_s$ is valid for the GBW model. By matching the asymptotic behavior of the saturation momentum at large $Y$ with the saturation momentum in the GBW model, we find that  $\alpha_s \approx 0.062$ which is unusually small value for the strong coupling constant. Such small value of the strong coupling constant is an artifact of the large higher order corrections to the BK equation. If the higher order corrections and possibly resummation  is included, the energy dependence extracted from the numerical solution of the BK equation should be close to the measured value  with a  
reasonable value of $\alpha_s$ (see, e.g., in Ref.~\cite{Triantafyllopoulos:2002nz}).

With the fixed coupling, it seems that the BK evolution produces a transient spike in $F_{x_g}$ at low
$k^2$, which is primarily due to our choice of the initial
condition which differs substantially from the asymptotic solution
at low $k^2$. Given the fact that our code samples $F_{x_g}$ within some extended region of $q^2_\perp$ for a certain value of $\Delta \phi$, this has very limited effect on
our results.

\subsubsection{Running Coupling}

Incorporating running coupling corrections into the leading order BK evolution is known to give a reasonably accurate description of DIS data~\cite{Albacete:2009fh,Albacete:2010sy}, and also to slow down the energy dependence of the saturation scale, which makes it a viable and perhaps preferable alternative to artificially tuning the value of $\alpha_s$. In this case, the saturation scale dependence has also been calculated in~\cite{Munier:2003vc} to be 
\begin{equation}\label{eq:satscalerunning}
 Q_s^2(Y) \propto \Lambda_\text{QCD}^2\exp\Biggl(\sqrt{\frac{2\chi(\gamma_0)}{\beta(1 - \gamma_0)}Y} + \frac{3}{4}\biggl(\frac{\chi''(\gamma_0)}{\sqrt{2\beta(1-\gamma_0)\chi(\gamma_0)}}\biggr)^{\frac{1}{3}}\xi_1 Y^{\frac{1}{6}}\Biggr)
\end{equation}
with a running coupling of the form
\begin{equation}
 \alphasbar(k^2) = \frac{1}{\beta\ln k^2/\Lambda_\text{QCD}^2},
\end{equation}
where $\beta = \frac{11 - 2N_f/N_c}{12}$, $\xi_1 = -2.338\ldots$ is the rightmost zero of the Airy function, $N_f = 3$ is the number of quark flavors that are expected to contribute, and $N_c = 3$ is the number of colors as before. Ref.~\cite{Mueller:2002zm} also includes a similar result.

However, this form is not practical for the numerical implementation due to a Landau pole at $k = \Lambda_\text{QCD}$. We use an implementation which shifts the squared momentum,
\begin{equation}\label{eq:runningcoupling}
 \alphasbar(k^2) = \frac{1}{\beta\ln\frac{k^2 + \mu^2}{\Lambda_\text{QCD}^2}}
\end{equation}
with $\mu = \SI{1}{GeV}$. The shift term $\mu^2$ is a purely phenomenologically motivated change, which does not appear to significantly alter the behavior of $Q_s(Y)$ discussed below.

In equation~\eqref{eq:runningcoupling}, we are treating $\Lambda_\text{QCD}^2$ as an adjustable parameter which absorbs the free parameter $4C^2$ of Ref.~\cite{Albacete:2009fh}. We can alter the behavior of the coupling, and thus the energy dependence of the solution, by adjusting the value of $\Lambda_\text{QCD}^2$. According to equation~\eqref{eq:satscalerunning}, the behavior of the saturation scale asymptotes to
\begin{equation}
 Q_s \to e^{\lambda_r\sqrt{Y}},\quad\text{where }\lambda_r = \sqrt{\frac{2\chi(\gamma_0)}{\beta(1 - \gamma_0)}}.
\end{equation}
This is qualitatively different from the large-$Y$ behavior of the GBW model, thus it prevents us from reproducing the asymptotic behavior of that model. Nevertheless, we can choose a value of $\Lambda_\text{QCD}$ that approximately matches $\pd{Q_s}{Y}$ to its value from the GBW model over the upper range of rapidities we calculate, $10\lesssim Y\lesssim 20$. Some experimentation shows that $\Lambda_\text{QCD}^2 = \SI{0.001}{GeV^2}$ gives a reasonable match. The corresponding value from the fits in~\cite{Albacete:2009fh,Albacete:2010sy} would be $\Lambda_\text{QCD}^2 \approx \frac{(\SI{0.241}{GeV})^2}{4(6)} = \SI{0.002}{GeV^2}$, which is fairly close.

Note that the value of the effective $\Lambda_\text{QCD}$ parameter determined by our fit is fairly small, which possibly indicates the need for further higher order corrections in the BK evolution (apart from the running coupling).

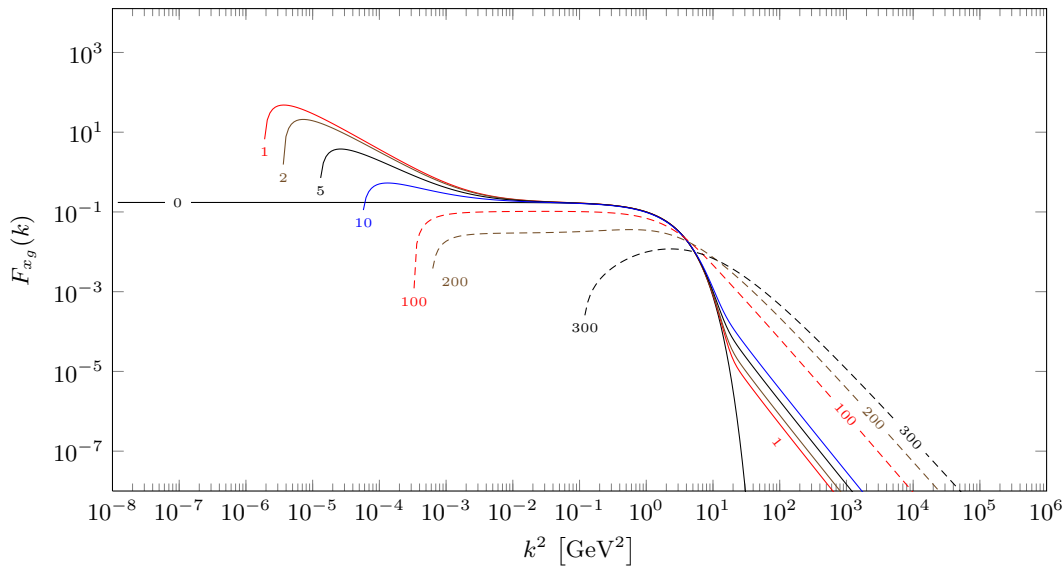
\begin{figure}
 % This figure may require at least PGF 2.10 to compile correctly.
 \begin{tikzpicture}
  \begin{axis}[yshift=-8cm,xmode=log,ymode=log,xmin=1e-8,xmax=1e6,ymin=1e-8,ymax=1e3,xlabel=$k^2$,ylabel=$F_{x_g}(k)$,x unit=GeV^2,width=14cm,height=8cm]
   \addplot+[black,no markers] file {datafiles/Fg/BK-Pb85-run-000.dat} node[pos=0.04,font=\tiny,fill=white] {$0$};
   \addplot+[no markers] file {datafiles/Fg/BK-Pb85-run-001.dat} node[pos=0,below,font=\tiny] {$1$} node[pos=0.5,sloped,below,font=\tiny] {$1$};
   \addplot+[no markers] file {datafiles/Fg/BK-Pb85-run-002.dat} node[pos=0,below,font=\tiny] {$2$};
   \addplot+[no markers] file {datafiles/Fg/BK-Pb85-run-005.dat} node[pos=0,below,font=\tiny] {$5$};
   \addplot+[no markers] file {datafiles/Fg/BK-Pb85-run-010.dat} node[pos=0,below,font=\tiny] {$10$};
   \addplot+[no markers] file {datafiles/Fg/BK-Pb85-run-100.dat} node[pos=0,below,font=\tiny] {$100$} node[pos=0.5,sloped,font=\tiny,fill=white] {$100$};
   \addplot+[no markers] file {datafiles/Fg/BK-Pb85-run-200.dat} node[pos=0,below right,font=\tiny] {$200$} node[pos=0.5,sloped,font=\tiny,fill=white] {$200$};
   \addplot+[no markers] file {datafiles/Fg/BK-Pb85-run-300.dat} node[pos=0,below,font=\tiny] {$300$} node[pos=0.5,sloped,font=\tiny,fill=white] {$300$};
  \end{axis}
 \end{tikzpicture}
 \caption{The gluon distribution from the numerical BK evolution with the running coupling for central \pPb{} collisions. This is a continuation of figure~\ref{fig:fgplot}, and the labels have the same interpretation. As is well known, the running coupling slows down the evolution of the wavefront. Otherwise, the plots are similar to the fixed-coupling BK evolution, though there is some erratic behavior at lower momenta especially at high rapidities which was not seen to such a large extent in the fixed coupling case.}
 \label{fig:fgplotrunning}
\end{figure}

The solution we have obtained for the BK evolution with running coupling is shown in figure~\ref{fig:fgplotrunning}. Although the overall shapes of the curves are qualitatively similar, the evolution with rapidity is slowed by the running coupling. Overall, the actual value of $F_{x_g}$ is less at large values of $k^2$ than it was in the fixed coupling case.

\section{Results}\label{sec:results}
\begin{figure}[p]
 \begin{tikzpicture}
  \begin{groupplot}[group style={rows=2,xlabels at=edge bottom,ylabels at=edge left},width=10cm,correlation graph,legend style={at={(0.95,0.97)},anchor=north east}]
   \nextgroupplot
   \addplot+[smooth] file {datafiles/rhic.252585.p12666.dat}; % RHIC GBW M=0.5
   \addplot+[smooth] file {datafiles/rhic.252585.p27336.dat}; % RHIC Mixed M=0.5
   \addplot+[smooth] file {datafiles/rhic.252585.p20033.dat}; % RHIC RC Mixed M=0.5
   \legend{GBW,BK,rcBK};
   \nextgroupplot
   \addplot+[smooth] file {datafiles/rhic.252585.p17509.dat}; % RHIC GBW M=4
   \addplot+[smooth] file {datafiles/rhic.252585.p27337.dat}; % RHIC Mixed M=4
   \addplot+[smooth] file {datafiles/rhic.252585.p30742.dat}; % RHIC RC Mixed M=4
   \legend{GBW,BK,rcBK};
  \end{groupplot}
 \end{tikzpicture}
 \caption{The angular correlations between the virtual photons and pions at RHIC, at medium rapidity, $Y_{\gamma} = Y_{\pi} = 2.5$. The upper graph shows the correlation for a virtual photon mass of $M = \SI{0.5}{GeV}$, and the lower one, for $M = \SI{4}{GeV}$. In each case, the three curves for GBW, fixed coupling BK, and running coupling BK, exhibit basically the same double-peak structure around $\Delta\phi = \pi$, but they show differing behavior near $\Delta\phi = 0,2\pi$, the near side correlation. This relates to the large-$k^2$ behavior of the corresponding gluon distributions.}
 \label{fig:corr_rhic_2}
\end{figure}
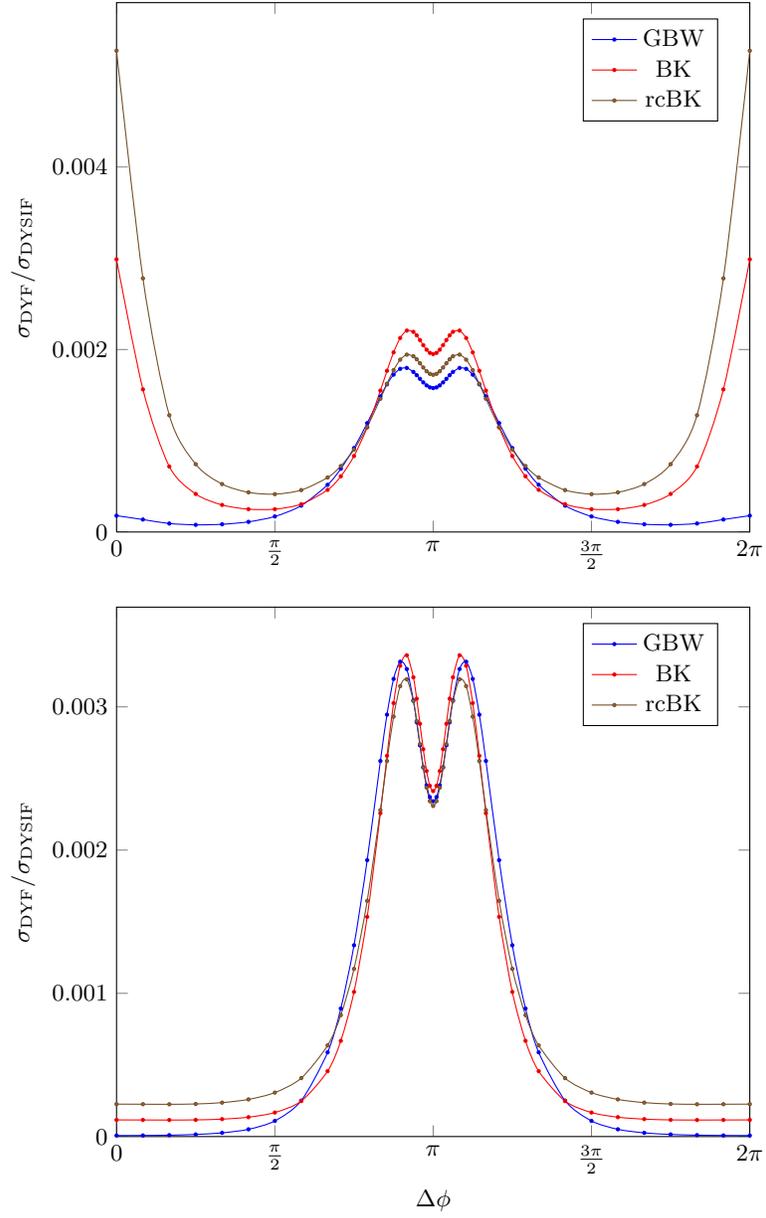
\begin{figure}[p]
 \begin{tikzpicture}
  \begin{groupplot}[group style={rows=2,xlabels at=edge bottom,ylabels at=edge left},width=10cm,correlation graph,legend style={at={(0.95,0.97)},anchor=north east}]
   \nextgroupplot
   \addplot+[smooth] file {datafiles/lhc.404085.p7033.dat};  % LHC GBW M=4
   \addplot+[smooth] file {datafiles/lhc.404085.p2584.dat};  % LHC Mixed M=4
   \addplot+[smooth] file {datafiles/lhc.404085.p28301.dat}; % LHC RC Mixed M=4
   \legend{GBW,BK,rcBK};
   \nextgroupplot
   \addplot+[smooth] file {datafiles/lhc.404085.p24755.dat}; % LHC GBW M=8
   \addplot+[smooth] file {datafiles/lhc.404085.p13089.dat}; % LHC Mixed M=8
   \addplot+[smooth] file {datafiles/lhc.404085.p6263.dat};  % LHC RC Mixed M=8
   \legend{GBW,BK,rcBK};
  \end{groupplot}
 \end{tikzpicture}
 \caption{The virtual photon-pion angular correlations at LHC at rapidity $Y_{\gamma} = Y_{\pi} = 4$. The upper and lower graphs show $M = \SI{4}{GeV}$ and $M = \SI{8}{GeV}$, respectively. As in figure~\ref{fig:corr_rhic_2}, the key difference is in the near side correlation, arising due to differences in the high-momentum region of the gluon distribution.}
 \label{fig:corr_lhc_2}
\end{figure}
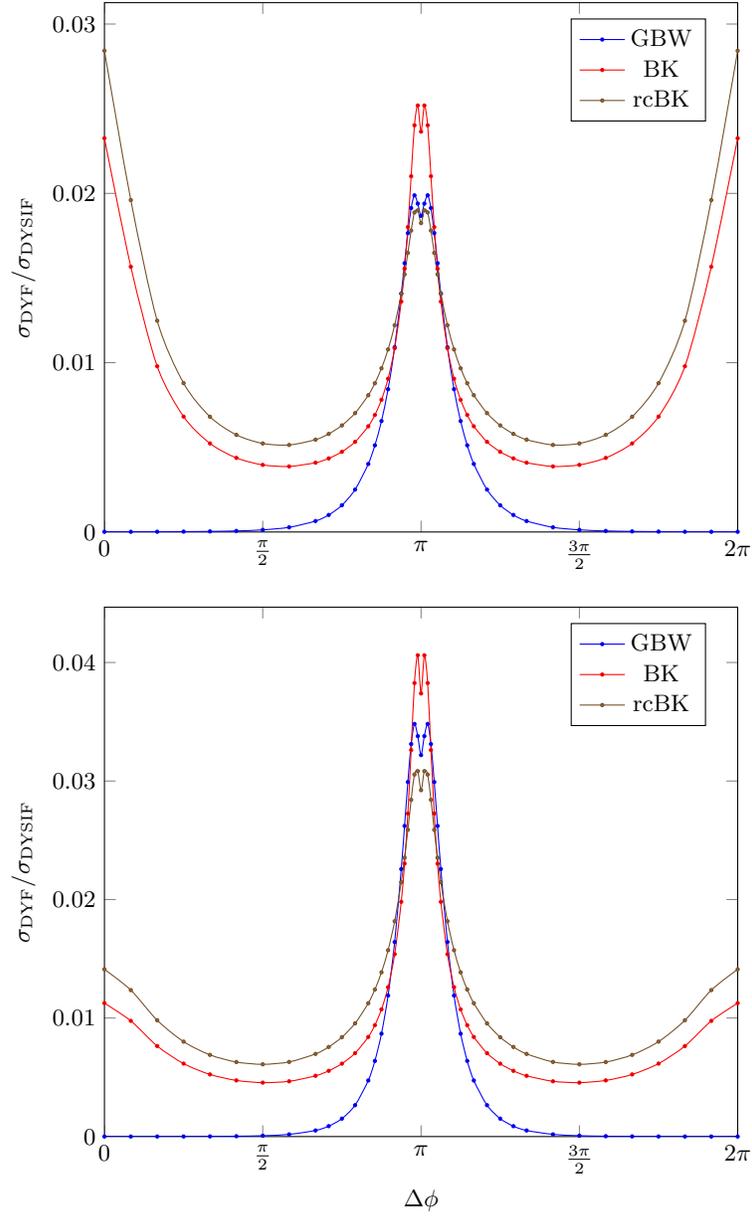

With the gluon distribution determined from either the analytic GBW model and the numerical solution of the BK evolution described in section \ref{sec:gluondist}, we can compute and study the correlation defined in section \ref{sec:dijetcorr}. There are eight parameters involved in this calculation:
\begin{itemize}
 \item the virtual photon mass $M$
 \item the photon and pion rapidities $Y_\gamma$ and $Y_\pi$
 \item the profile function $c(b)$
 \item the mass number $A$
 \item the center-of-mass energy $\sqrt{s}$ per nucleon
 \item the detector-imposed cutoff on the transverse momenta of the virtual photons or pions, $\ptcut$
 \item Different types of projectiles: either deuterons or protons.
\end{itemize}
To simulate results from \dAu{} collisions at RHIC, we set $\sqrt{s} = \SI{200}{GeV}$, $A = 197$, and $p_{\perp\text{cut}} = \SI{1.5}{GeV}$ together with the use of the deuteron parton distribution functions. For LHC \pPb{} collisions, we put $\sqrt{s} = \SI{8800}{GeV}$ which corresponds to the full design energy ($\SI{14}{TeV}$ pp CM energy), $A = 208$, and $p_{\perp\text{cut}} = \SI{3}{GeV}$. For both the RHIC and LHC measurement, we compute the correlation for a low and a high value of $M$. In addition, we set $Y_\gamma = Y_\pi$ which means that the virtual photons and pions are produced in the same rapidity window, and use $c(b) = 0.85$ which corresponds to central collisions.

We show our final results for the correlations in figures \ref{fig:corr_rhic_2} and \ref{fig:corr_lhc_2} with different specifically chosen parameters. From figs.~\ref{fig:corr_rhic_2} and \ref{fig:corr_lhc_2}, we find that indeed the GBW model and the BK equation with fixed and running couplings yield similar double-peak structure, as expected from our early discussion for the correlation function at the away side ($\Delta \phi \simeq \pi$). However, they differ dramatically at the near side ($\Delta \phi \simeq 0 \text{ or } 2\pi$). It is easy to see that the near side correlation samples the large $q_\perp$ region of the gluon distribution $F_{x_g}(q_\perp)$. Since this distribution is slightly greater with the running coupling than with fixed coupling, the near side peak is somewhat enhanced in the running coupling case.

We believe that the solution to the BK equation gives the correct large $q_\perp$ behavior for the dipole gluon distribution while the GBW model fails at large $q_\perp$. Therefore, we expect that the curves generated by the numeric solution of the BK equation predicts the correlation for RHIC and LHC measurement for all azimuthal angle including both the near side and the away side.

\section{Conclusion}\label{sec:conclusion}
In this paper, by using the GBW model and the numerical solution
to the BK equation, we have numerically studied the correlations
between the lepton-pair and associated hadron in Drell-Yan processes
in \pA{} collisions, which directly probe the dipole gluon
distributions. The correlations have been computed for both RHIC
\dAu{} collisions and LHC \pPb{} collisions. We find that the Drell-Yan type correlation implies a
strong suppression of the away side hadron at forward rapidity due
to strong interaction between the forward quark from the
projectile proton and the dense gluon from the target nucleus. In
addition, we also emphasize that the Drell-Yan type correlation
exhibits a unique double-peak structure in the away side.

\begin{acknowledgments}
We thank E. Avsar, F. Dominguez, A. Mueller and F. Yuan for helpful conversations. This work was
supported in part by the U.S. Department of Energy under the
contract DOE OJI grant No. DE - SC0002145 and by the Polish NCN
grant DEC-2011/01/B/ST2/03915.
AMS is supported by the Sloan Foundation.
We are grateful to RIKEN, Brookhaven National Laboratory and the
U.S. Department of Energy (contract number DE-AC02-98CH10886) for
providing the facilities essential for the completion of this
work.
\end{acknowledgments}

\end{document}